\documentclass[reprint, twocolumn, secnumarabic,amssymb, aps, prb]{revtex4-1}
\usepackage{amsmath}
\usepackage{graphicx}

\begin{document}

\title{Layer-number determination in graphene by out-of-plane phonons}%

\author{Felix Herziger}%
\email{fhz@phyik.tu-berlin.de}

\author{Patrick May}%

\author{Janina Maultzsch}%
\affiliation{Institut f\"ur Festk\"orperphysik, Technische Universit\"at Berlin, Hardenbergstr. 36, 10623 Berlin, Germany}

\date{\today}%

\begin{abstract}
We present and discuss a double-resonant Raman mode in few-layer graphene, which has not been interpreted before and is able to probe the number of graphene layers. This so-called $N$ mode on the low-frequency side of the $G$ mode results from a double-resonant Stokes/anti-Stokes process combining an optical (LO) and an out-of-plane (ZO$'$) phonon. Simulations of the double-resonant Raman spectra in bilayer graphene show very good agreement with the experiments. 
\end{abstract}

\maketitle

Raman spectroscopy belongs to the most widely used methods in graphene research. Raman spectroscopy is used for characterizing graphene regarding defects \cite{10.1021/nl201432g,10.1103/PhysRevB.82.125429,10.1103/PhysRevB.84.035433}, doping \cite{10.1038/nmat1846,10.1038/nnano.2008.67}, strain \cite{10.1103/PhysRevB.79.205433,10.1103/PhysRevB.82.201409,10.1021/nn800459e,10.1103/PhysRevB.85.115451}, crystallographic orientation \cite{10.1073/pnas.0811754106,10.1021/nl8032697}, or interaction with the substrate \cite{10.1021/jp8008404}. In view of fundamental physical properties of graphene, Raman spectroscopy gives information on electron-phonon coupling and scattering rates, optical excitations in graphene, thermal and mechanical properties \cite{10.1103/PhysRevLett.93.185503, 10.1021/nl801884n, 10.1103/PhysRevB.75.045404}. Probably the most popular application of Raman scattering in graphene is the distinction of single-layer graphene from few-layer graphene and graphite via the lineshape of the double-resonant $2D$ mode \cite{10.1103/PhysRevLett.97.187401}. On the other hand, few-layer graphene has recently come into focus, as gated bi- and trilayer graphene offer a tunable band gap \cite{10.1038/nature08105, 10.1038/nphys2102} and bilayer graphene has been demonstrated to give much higher on-off ratios in a field-effect transistor than single-layer graphene \cite{10.1021/nl9039636}. Therefore, it is important to establish a reliable method for the determination of the layer number in few-layer graphene and to identify spectroscopic signatures of the layer-layer interaction. So far, typically the evolution of the $2D$-mode lineshape or the absolute Raman intensity of the $G$ mode is used in combination with optical contrast measurements. However, the lineshape of the $2D$ mode depends strongly on the excitation wavelength \cite{10.1103/PhysRevLett.97.187401}, and the $G$-mode amplitude depends not only on the scattering volume \cite{10.1021/nl061420a,10.1021/nl061702a}, but also on the substrate and optical interference effects \cite{10.1103/PhysRevB.80.125422}. Recently, the rigid-layer shear mode, which is the other Raman-active $E_{2g}$ phonon mode in graphite, was shown to have a strong frequency dependence on the number of layers in few-layer graphene \cite{10.1038/nmat3245}. The frequency of this mode, however, is below 44\,cm$^{-1}$. Measurement of this low-frequency mode is therefore difficult and requires non-standard equipment.

Here we present and interpret a newly discovered Raman mode on the low-frequency side of the $G$ mode, which can be used to determine the number of layers in few-layer graphene. This so-called $N$ mode is based on a double-resonant intravalley scattering process combining the longitudinal optical (LO) and the rigid-layer compression mode (ZO$'$). The peak position as well as the lineshape of this peak allow an assignment of the Raman spectra to the number of graphene layers for up to approximately eight layers. In addition, we simulate the double-resonant Raman spectra in the $N$-mode region for various excitation energies in bilayer graphene.

\begin{figure}[t]
\includegraphics[width=\columnwidth]{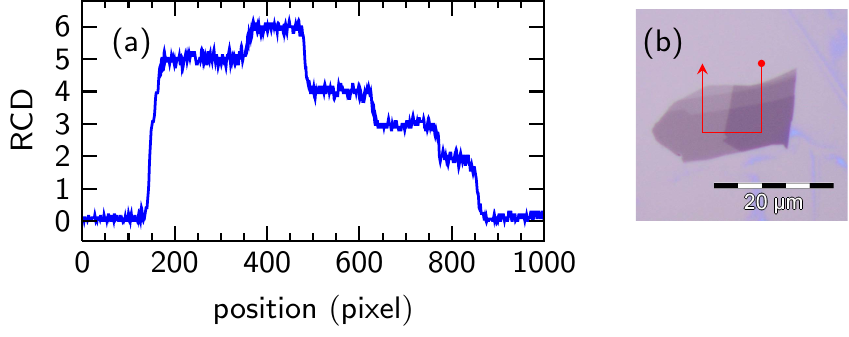}
\caption{(a) Ratio of Color Difference (RCD) measurement on a few-layer graphene sample. (b) Few-layer graphene and the scanning path on the sample. Lengths along the RCD scan are not drawn to scale. \label{fig:RCD}}
\end{figure}

Graphene samples were prepared by mechanical cleavage under cleanroom conditions from natural graphite flakes and transferred onto a silicon substrate with an oxide thickness of 80\,nm. The samples were analyzed with an optical microscope (Olympus BX51M with an 100$\times$ objective). We determined the number of graphene layers by optical contrast, using the Ratio of Color Difference (RCD) method \cite{10.1021/nl071254m}. The RCD values were calculated using the formalism from Ref.[\onlinecite{10.1021/jp1121596}] 
\begin{equation}
\text{RCD} = \sqrt{\sum\limits_{i=X,Y,Z}\left(i_n - i_0\right)^2}\bigg/\sqrt{\sum\limits_{i=X,Y,Z}\left(i_1 - i_0\right)^2}
\end{equation}
where $X_0$, $Y_0$, and $Z_0$ denote the tristimulus color components of the Si/SiO$_2$ substrate and $X_n$, $Y_n$, and $Z_n$ the color components of \textit{n}-layer graphene. Since the RCD is independent of the light source \cite{10.1021/jp1121596}, the RCD values can be calculated directly from the RGB color values of the optical image. In Fig.\,\ref{fig:RCD}(a) an exemplary result of a RCD scan is shown. Here, the RCD measurement along the highlighted path reveiled graphene thicknesses ranging from \textit{n}=2 to \textit{n}=6 layers. This result corresponds to the optical contrast from the image, which is shown in Fig.\,\ref{fig:RCD}(b). Graphene samples with layer thicknesses up to 11 layers have been prepared and were characterized by this method.

We performed confocal $\mu$-Raman measurements under ambient conditions using a LabRAM HR800 spectrometer. Laser excitation wavelengths of 532\,nm (2.33\,eV) and 633\,nm (1.96\,eV) were chosen. Raman spectra were recorded in back-scattering geometry with a spectral resolution better than 1\,cm$^{-1}$. The laser was focussed with an 100$\times$ objective and had a spot size $<$500\,nm. All spectra were calibrated by standard atomic emission lines of neon (Ne).

The bandstructure and phonon dispersion of bilayer graphene were calculated using the \textsc{Siesta} DFT code in local-density approximation \cite{10.1088/0953-8984/14/11/302}. The calculations were performed according to Ref.\,\onlinecite{10.1103/PhysRevB.81.205426}. We used the experimental geometrical values of graphite, \textit{i.e.}, a lattice constant of $a=2.46$\,\AA{} and an interlayer distance of $c/2 = 3.35$\,\AA{} \cite{10.1103/PhysRevB.76.035439}. The $\Gamma$-point frequency of the $E_{\text{2g}}$ mode was scaled by a factor of 0.96 to the experimental value of 1584\,cm$^{-1}$. We rescaled the calculated phonon dispersion by the same factor; the resulting phonon dispersion shows very good agreement with experimental data \cite{10.1103/PhysRevB.76.035439}.

\begin{figure}[t]
\includegraphics[width=\columnwidth]{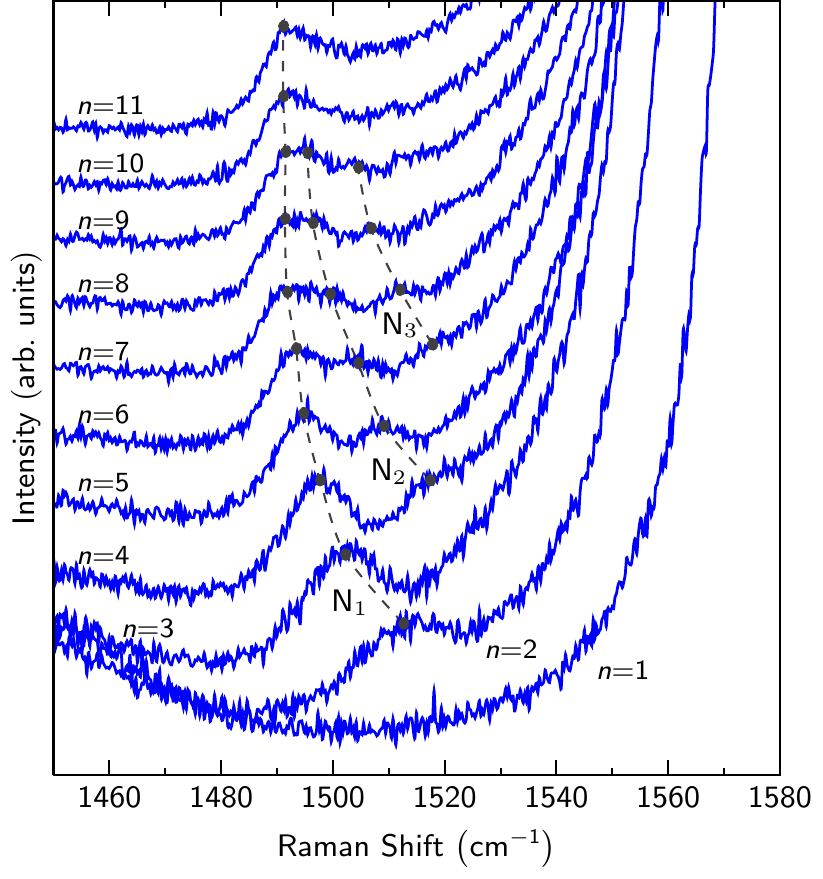}
\caption{Raman spectra of \textit{n}-layer graphene on the low-frequency side of the $G$ mode at 633\,nm laser wavelength. Spectra are normalized to the same $G$-mode amplitude and vertically offset for clarity. The dashed lines serve as a guide to the eye. \label{fig:RamanZO-}}
\end{figure}

\begin{figure}[t]
\includegraphics[width=\columnwidth]{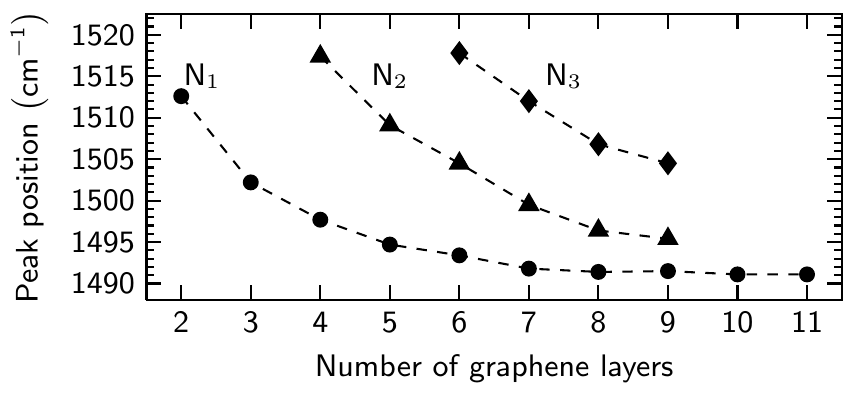}
\caption{Peak positions of the $N$ mode as a function of the number of graphene layers at 633\,nm laser wavelength. \label{fig:peaks}}
\end{figure}

Fig.\,\ref{fig:RamanZO-} shows the Raman spectra of \textit{n}-layer graphene for layer thicknesses ranging from monolayer to 11-layer graphene at 633\,nm laser wavelength. For $n \geq 2$ we observe a layer-dependent peak on the low-frequency side of the $G$ mode. This mode is approximately 100 times weaker than the $G$ mode. It is clearly absent in monolayer graphene. Furthermore, additional peaks appear for more layers. We label these Raman modes in the order of their appearance as $N_1$, $N_2$, and $N_3$ and refer to them as $N$ mode. The layer-dependent shift of their peak position is shown in Fig.\,\ref{fig:peaks}. The frequencies decrease and tend towards a lower limit as the layer thickness is increased. 


\begin{figure}[t]
\includegraphics[width=\columnwidth]{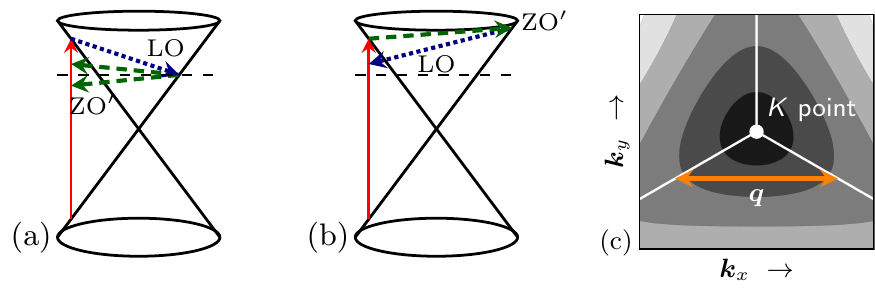}
\caption{Schematic illustration of the double-resonant intravalley scattering of a LO phonon (blue, dotted) and a ZO$'$ phonon (green, dashed). Energies are not drawn to scale. In (a) the electron is first scattered by the LO phonon, then the Stokes/anti-Stokes scattering with a ZO$'$ phonon follows. (b) shows the reversed order. (c) Contour plot of graphene's bandstructure around a $K$ point. The $K$-$M$ direction is highlighted in white. The orange arrow denotes the phonon wave vector $q$ that is enhanced by the double resonance. \label{fig:RamanDR}}
\end{figure}

The absence of the $N$ mode in single layer graphene indicates that it may originate from interlayer vibrations. We assign the $N$ mode to a double-resonant intravalley scattering close to the $K$ point combining LO (longitudinal optical) and ZO$'$ (rigid-layer compression) phonons, in which the LO phonon is Stokes- and the ZO$'$ phonon anti-Stokes scattered. An illustration of the double resonance is shown in Fig.\,\ref{fig:RamanDR}. The dashed horizontal line corresponds to the phonon frequency of the defect-scattered LO phonon, \textit{i.e.}, the $D'$ mode. In (a) the electron is first scattered by an LO phonon and afterwards the Stokes/anti-Stokes scattering with a ZO$'$ phonon follows. The reversed order in the scattering process is shown in Fig.\,\ref{fig:RamanDR}(b). The double-resonant scattering in a two-dimensional illustration is shown in Fig.\,\ref{fig:RamanDR}(c). The resonantly enhanced phonon wave vector along $\Gamma$-$M$ connects two electronic states on the $K$-$M$ high-symmetry line. An explanation of this scattering process is given below.

\begin{figure}[t]
\includegraphics[width=\columnwidth]{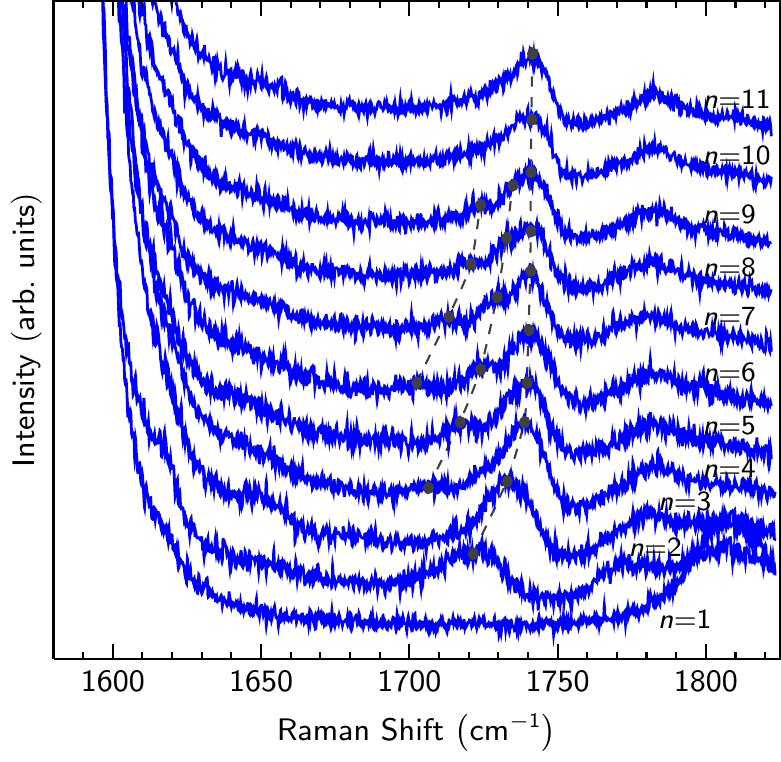}
\caption{Raman spectra of the double-resonant LOZO$'^+$ peak in \textit{n}-layer graphene at 633\,nm laser excitation wavelength. Spectra are normalized and vertically offset for clarity. The dashed lines serve as a guide to the eye. \label{fig:RamanZO+}}
\end{figure}

Our assumption is supported by the correspondence between the $N$ mode and the double-resonant LO+ZO$'$ peak ($\sim$1740\,cm$^{-1}$), resulting from an intravalley double resonance combining an LO and ZO$'$ phonon \cite{10.1021/nn203472f,10.1021/nn1031017}. We label this peak in the following as LOZO$'^+$. The Raman spectra of the LOZO$'^+$ peak for layer thicknesses from monolayer to 11-layer graphene are shown in Fig.\,\ref{fig:RamanZO+}. All peaks of the $N$ mode and the LOZO$'^+$ peak have approximately the same distance to the $D'$ mode, which can be resolved at $\sim$1616\,cm$^{-1}$ for 633\,nm laser wavelength. Due to this symmetry, both the $N$ mode and LOZO$'^+$ peak must differ from the D$'$ mode in the same process, namely by the scattering with a ZO$'$ phonon. In the case of the $N$ mode, the ZO$'$ phonon is anti-Stokes scattered, whereas the ZO$'$ phonon is Stokes-scattered for the LOZO$'^+$ peak. This combination of Stokes and anti-Stokes scattered phonons in a double-resonant process was never reported before.

\begin{figure}
\includegraphics[width=\columnwidth]{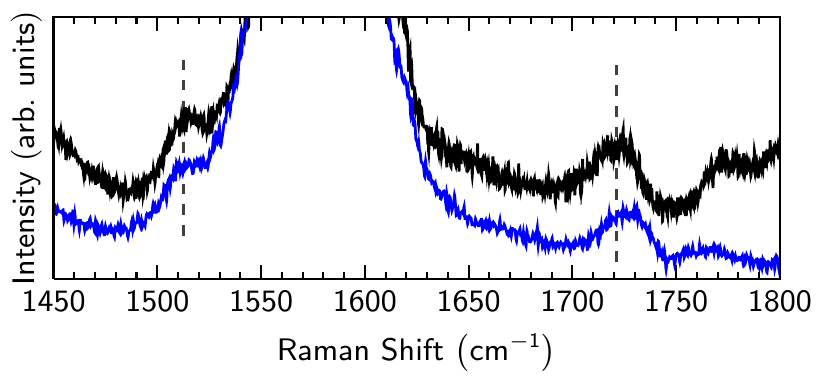}
\caption{Comparison of the $N$ mode and LOZO$^+$ peak in bilayer graphene for 532\,nm (blue, lower spectrum) and 633\,nm (black, upper spectrum) laser excitation wavelength. \label{fig:Disp}}
\end{figure}

\begin{figure}
\includegraphics[width=\columnwidth]{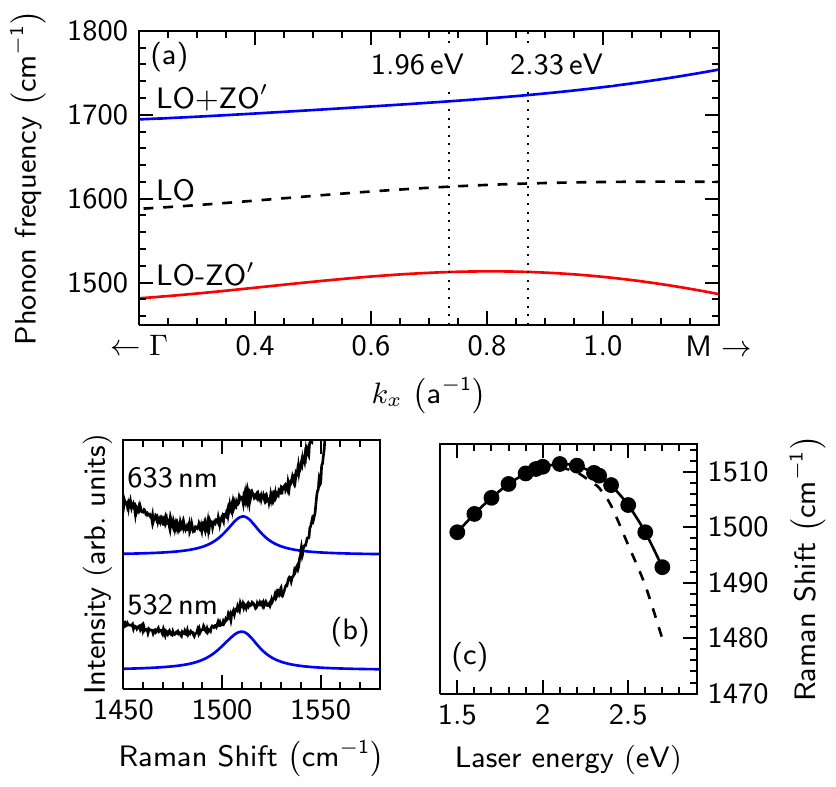}
\caption{(a) LO phonon branch (dashed), LO-ZO phonon branch (lower curve), and the LO+ZO phonon branch (upper curve) along the $\Gamma$-M direction in bilayer graphene. The M point corresponds to $2\,\pi/(\sqrt{3}\,a)$ with $a=2.46\,\text{\AA}$. The vertical dotted lines denote the phonon wavevector at 1.96\,eV and 2.33\,eV laser excitation energy, respectively. (b) Simulated double-resonant Raman spectra in bilayer graphene (blue curve) in the $N$ mode region compared to experimental spectra (black curve). (c) Laser energy dependent peak positions of the $N$ mode calculated for bilayer graphene.  \label{fig:LOZOcalc}}
\end{figure}

Since we assign the $N$ mode to a double-resonant Raman process, we would expect a laser energy dependent shift of the peak position, as this is a fingerprint of double-resonant Raman scattering. Fig.\,\ref{fig:Disp} shows the spectra of bilayer graphene for 532\,nm and 633\,nm excitation wavelength. The LOZO$'^+$ peak blueshifts with increasing laser wavelength, whereas a shift of the $N$ mode cannot be observed or is on the order of our spectral resolution. This behavior can be understood from the dispersion of the LO and ZO$'$ phonon branch shown in Fig.\,\ref{fig:LOZOcalc}(a). When a double-resonant process combines Stokes and anti-Stokes scattered phonons, the difference of both phonon frequencies determines the final peak position. The LO and ZO$'$ phonon branch exhibit nearly the same slope in the relevant range. Hence, the difference of both phonon branches is nearly constant. Therefore, a change of the phonon wavevector does not result in a shift of the phonon frequency and the $N$ mode shows no or little dispersion in the range between 1.9\,eV and 2.3\,eV laser energy. In fact, the shift of the $N$ mode between 633\,nm and 532\,nm excitation wavelength, derived from the phonon dispersion in Fig.\,\ref{fig:LOZOcalc}(a), is less than 1\,cm$^{-1}$. This result fits our experimental observations very well. For the LOZO$'^+$ peak, both phonon branches must be added. The resulting phonon branch has a positive slope; therefore the peak position should increase for higher excitation energies. We estimate from Fig.\,\ref{fig:LOZOcalc}(a) a blueshift of $\sim$7\,cm$^{-1}$, which is close to the experimentally obtained shift of +5\,cm$^{-1}$. 

The shift of the $N_i$ peaks ($i=1,2,3$) as a function of the number of graphene layers can be explained with the evolution of the ZO$'$ phonon spectra in few-layer graphene. In \textit{n}-layer graphene there exist \textit{n}-1 vibrations with rigid-layer compression pattern \cite{10.1103/PhysRevB.78.085424, 10.1103/PhysRevB.85.094303}. Therefore, for an increasing number of graphene layers, the LO phonon can scatter with an increasing number of ZO$'$ phonons. The ZO$'$ vibrations exhibit a layer-dependent shift towards an upper limit, \textit{i.e.}, the frequency in bulk graphite \cite{10.1103/PhysRevB.78.085424, 10.1103/PhysRevB.85.094303}. This explains the downshift of the $N$ mode and the upshift of the LOZO$'^+$ peak as a function of the number of graphene layers, as well as the appearance of additional peaks for increasing number of layers. Recent work from Lui \textit{et al.} show a similar layer-dependence of the LOZO$'^+$ mode on the high-frequency side of the $G$ mode for up to six layers \cite{arXiv:1204.1702}, in agreement with the spectra shown in Fig.\,\ref{fig:RamanZO+}.

To support our interpretation, we simulated the double-resonant Raman spectra using the equation \cite{10.1103/PhysRevLett.85.5214}
\begin{align}
\left\vert K_{2f,10}\right\vert^2 = \bigg\vert&\sum\limits_{a,b,c}\frac{\mathcal{M}}{\left(E_L - E_{ai}-i\gamma\right)\left(E_L - E_{bi}-\hbar\omega_{LO} - i\gamma\right)} \notag \\
\times&\frac{1}{\left(E_L - E_{ci} -\hbar\omega_{LO} + \hbar\omega_{ZO'} - i\gamma\right)} \bigg\vert^2 
\end{align}
where $E_L$ is the energy of the incoming photon and $\mathcal{M}$ are the matrix elements, which are assumed to be constant. However, the strong angular dependence of the optical matrix elements was taken into account by setting the integration path as shown in Fig.\,\ref{fig:RamanDR}(c), in agreement with results for the $D'$ mode in Ref.\,\cite{10.1103/PhysRevB.84.035433}. Thus, the optical transitions are calculated along $K$-$M$, whereas the phonons predominantly stem from the $\Gamma$-$M$ direction. The energy differences between the intermediate electronic states $a$, $b$, $c$ and the initial state $i$ are labeled as $E_{xi}$. The broadening factor $\gamma$ was set to 40\,meV \cite{10.1103/PhysRevB.84.035433}. Our calculations also include the reversed order, where the ZO$'$ phonon is scattered first and the LO phonon secondly, and scattering by both electrons and holes.

Results of our calculations for bilayer graphene are shown in Fig.\,\ref{fig:LOZOcalc}(b) and (c). The simulated spectra in Fig.\,\ref{fig:LOZOcalc}(b) fit our experimental data very well. The laser-energy dependent peak position of the $N$ mode is shown in Fig.\,\ref{fig:LOZOcalc}(c). The $N$ mode follows the dispersion of the LO-ZO$'$ phonon branch. The laser-dependent peak shift is in the visible range much less than that of the LOZO$^+$ peak, in agreement with the experiments. At higher excitation energies above $\sim$2.5\,eV, we observe a splitting of the $N$ mode due to distinct contributions from the two $\pi$ bands in bilayer graphene.

In summary, we have presented and interpreted a layer-number dependent Raman mode on the low-frequency side of the $G$ mode in few-layer graphene. This so-called $N$ mode is a combination mode of a Stokes-scattered LO phonon and an anti-Stokes scattered ZO$'$ phonon. The investigation of the peak positions enables determination of the number $n$ of graphene layers up to $n=8$.

The simulation of the double-resonant Raman spectra agrees very well with the experimental results. The $N$ mode shows in the visible range only little dispersion with laser wavelength. Furthermore, the $N$ mode does not overlap with other overtones or combinational modes, in contrast to the LO+ZO$'$ peak. Depending on the excitation wavelength, the $N$ mode may also be indicative of the stacking order in few-layer graphene. Furthermore, the study of ZO$'$ phonons can give information about the strenght of layer-layer interaction in few-layer graphene. 

Since the occurence of the ZO$'$ vibration is not restricted to graphene, this approach of determining the number of layers might be transferable to other layered materials.

\appendix
\begin{acknowledgments}
We thank the Fraunhofer IZM Berlin for the supply of substrates. This work was supported by the European Research Council (ERC) under grant no. 259286 and by the DFG under grant number MA 4079/3-1. 
\end{acknowledgments}

\bibliographystyle{apsrev4-1}

\end{document}